\documentclass[runningheads]{llncs}
\usepackage{amsmath}
\usepackage{amsfonts}
\usepackage{longtable}
\usepackage{graphicx}
\usepackage{adjustbox}
\usepackage{amssymb}
\usepackage{tabularx}
\usepackage{float}

\usepackage[T1]{fontenc}
\usepackage{graphicx}
\begin{document}
\title{Adaptive Extensions of Unbiased Risk Estimators for Unsupervised Magnetic Resonance Image Denoising}
\author{Reeshad Khan\inst{1}\orcidID{0009-0008-9870-022X} \and
Dr. John Gauch\inst{2}\orcidID{} \and
Dr. Ukash Nakarmi\inst{3}\orcidID{0000-0002-5351-3956}}
\authorrunning{F. Author et al.}
%
\institute{
    University of Arkansas, Fayetteville AR 72701, USA\\
    \email{jgauch@uark.edu}\\
    \and
    University of Arkansas, Fayetteville AR 72701, USA\\
    \email{rk010@uark.edu}\\
    \and
    University of Arkansas, Fayetteville AR 72701, USA\\
    \email{unakarmi@uark.edu}
}

\maketitle              
\begin{abstract}
The application of Deep Neural Networks (DNNs) to image denoising has notably challenged traditional denoising methods, particularly within complex noise scenarios prevalent in medical imaging. Despite the effectiveness of traditional and some DNN-based methods, their reliance on high-quality, noiseless ground truth images limits their practical utility. In response to this, our work introduces and benchmarks innovative unsupervised learning strategies, notably Stein's Unbiased Risk Estimator (SURE), its extension (eSURE), and our novel implementation, the Extended Poisson Unbiased Risk Estimator (ePURE), within medical imaging frameworks.
This paper presents a comprehensive evaluation of these methods on MRI data afflicted with Gaussian and Poisson noise types, a scenario typical in medical imaging but challenging for most denoising algorithms. Our main contribution lies in the effective adaptation and implementation of the SURE, eSURE, and particularly the ePURE frameworks for medical images, showcasing their robustness and efficacy in environments where traditional noiseless ground truth cannot be obtained.

\keywords{Deep Learning \and Stein's Unbiased Risk Estimator \and Extended Poisson Unbiased Risk Estimator \and MRI Denoising \and Unsupervised Learning.}

\end{abstract}
\section{Introduction}

Magnetic Resonance images(MRI) is a popular tool in clinical settings due to its non-ionizing properties and high-quality tissue contrast, without exposing the patient to radiation  \cite{IK2024Learn}. Despite its benefits, these images are inherently noisy due to its acquisition process
and thus filtering methods are usually required to improve their quality\cite{JV2019MRI}. In the rise of Deep learning, traditional denoising techniques, such as BM3D\cite{Har2012ID}, have been challenged by deep neural networks(DNNs) especially when dealing with synthetic Gaussian\cite{Kai2017DCNN} or Poisson noise\cite{SR2012Phy}. The evolution of DNNs showcases their superior capability in managing both synthetic and realistic noise scenarios. The standard training protocols for DNN-based denoisers typically require pristine, noiseless ground truth images. However, obtaining such clean images is often infeasible, prompting the need for alternative training strategies\cite{zhussip2019extending}. Historical efforts in creating benchmark datasets for denoising, like those by Plotz and Roth, involve averaging multiple noisy captures to simulate a single clean image, a method that becomes particularly complex with transient or subtle subject details as seen in medical or aerial imaging\cite{zhussip2019extending}.
This necessity has led to the emergence of unsupervised training techniques for DNNs. Methods like Deep Image Prior (DIP) leverage the intrinsic capabilities of generator networks to minimize the mean squared error (MSE) between noisy inputs and their denoised outputs\cite{Dim2018DIP}. Similarly, Noise2Noise, which trains networks using pairs of independently noisy images, has demonstrated impressive results in various applications, including MRI recovery without requiring clean ground truth data\cite{Jaak2018N2N}.Building upon these developments, Stein’s Unbiased Risk Estimator (SURE) and its extensions, such as eSURE, offer frameworks for training denoisers using single or correlated pairs of noisy images. This approach has been adapted further to develop the Poisson Unbiased Risk Estimator (PURE)\cite{Kim2020Un} and its enhanced version, Extended Poisson Unbiased Risk Estimator (ePURE)\cite{kim2022epure}, which we introduce in this work for MRI denoising. By employing correlated pairs of noisy images, ePURE facilitates the training of DNNs for Poisson noise scenarios, common in medical imaging, where traditional clean images are unavailable.
Our contribution, following the methodology from Magauiya et al. \cite{zhussip2019extending} and extending the ideas in Hanvit et al. \cite{kim2022epure}, focuses on applying ePURE\cite{kim2022epure} to train DNNs under conditions typical for Gaussian and Poisson noise contamination. This method not only proves to be superior to traditional SURE and eSURE\cite{zhussip2019extending} for Gaussian noise but also achieves competitive results for Poisson and mixed noise types, especially in the presence of imperfect ground truth scenarios. This paper provides a comprehensive theoretical and empirical analysis of the Noise 2 Noise\cite{Jaak2018N2N} approach adapted for correlated noise conditions and explores the advancements with eSURE and ePURE in handling various noise models. We conclude by reinforcing the theoretical connections between Noise2Noise and eSURE, positioning Noise2Noise as a special case of the eSURE framework for independent Gaussian noise scenarios. The structured discussions in this paper include a detailed review of the SURE methodology, extensions to eSURE and ePURE, and a critical evaluation of their effectiveness through experimental validations.

\section{Background}


\subsection{Stein‘s Unbiased Risk Estimator (SURE)}

In the context of denoising Gaussian-contaminated signals or images, a typical model involves a linear equation:
$$
y = x + n \eqno{(1)}
$$
where $x \in R^N$ represents an unknown signal, $y \in \mathbb{R}^N$ is a known measurement, and $n \in \mathbb{R}^N$ denotes i.i.d. Gaussian noise with $n \sim \mathcal{N}(0, \sigma ^2 I)$, where $I$ is an identity matrix. We express $n \sim \mathcal{N}(0, \sigma ^2 I)$ as $n \sim N_0,\sigma ^2$.
The SURE (Stein's Unbiased Risk Estimator) is a widely-used approach to estimate the mean squared error (MSE) associated with an estimator $h(y)$ of $x$. It is given by the following expression:
$$
\eta(h(y)) = \frac{||y - h(y)||^2}{N} - \sigma^2 + \frac{2 \sigma^2}{N} \sum_{i=1}^{N} \frac{\delta h_i(y)}{\delta y_i} \eqno{(2)}
$$
Assuming $x$ is a deterministic signal (or image),(2) establishes that the random variable $\eta(h(y))$ is an unbiased estimator of the $MSE(h(y))$\cite{CM1981}\cite{TB2007}, which is equivalent to:
$$
\mathbb{E}_{n \sim \mathcal{N}_{0,\sigma ^2}} \Biggl\{\frac{||x - h(y)||^2}{N}\Biggl\} = \mathbb{E}_{n \sim \mathcal{N}_{0,\sigma ^2}} \{\eta (h(y))\} \eqno{(3)}
$$
Here, $\mathbb{E}_{n \sim \mathcal{N}_{0,\sigma ^2}} \{ \cdot \}$ denotes the expectation operator in terms of the random vector n.

While the expression (2) is attractive for optimizing the parameters of an estimator h(y), obtaining an analytical solution for the last divergence term in (2) is limited to special cases, such as the estimator h(y) being a non-local mean or linear filter\cite{Dim2009} \cite{Dim2011}. Consequently, to utilize (2) in more general cases, it becomes necessary to find at least an approximate solution for the divergence term.

\subsection{Monte-Carlo SURE (MC-SURE)}

MC-SURE is a Monte Carlo method proposed by Ramani et al. \cite{Ram2008} to estimate the divergence and, consequently, the SURE loss. \\
Assume $\tilde{b} \sim \mathcal{N}_{0,1} \in \mathbb{R}^N$ is a Gaussian vector which is independent of $n$ or $y$. Ramani et al\cite{Ram2008} show that,

$$
\sum_{i=1}^{K} \frac{\delta h_i(y)}{\delta y_i} = \lim_{\epsilon \rightarrow 0} \mathbb{E}_{\tilde{b}}  \Biggl\{ \tilde{b}^t \Biggl( \frac{h(y+\epsilon \tilde{b}) - h(y)}{\epsilon} \Biggl) \Biggl\} \eqno{(4)}
$$
 Therefore, by applying this eq 4 to the divergence term in eq 2:
$$
\frac{1}{N} \sum_{i=1}^{N} \frac{\delta h_i(y)}{\delta y_i} \approx \frac{1}{\epsilon N} \tilde{b}^T (h(y+ \epsilon \tilde{b}) - h(y)) \eqno{(5)}
$$
Here, $\tilde{b}^T$ is the transpose of $\tilde{b}$, and $\epsilon$ is a small positive value to approximate the limit.

\subsection{Optimization of Deep Denoisers via SURE-Derived Losses}

The employment of Stein's Unbiased Risk Estimator (SURE) as an ersatz metric for the unsupervised optimization of deep neural network (DNN)-based denoisers has been gaining traction\cite{sha2018}. This method eschews the need for pristine ground truth data, which is a common limitation in supervised learning paradigms. Specifically, the Monte-Carlo SURE (MC-SURE) adaptation facilitates the training of DNN architectures by leveraging only noisy observations to approximate mean squared error (MSE) minimization between the denoised outputs and their ground truth counterparts. The reformulation of the MC-SURE objective for deep denoising tasks is articulated through the following equation, applied to a generic deep learning model parameterized by \(\theta\):
\begin{align}
\eta(h_{\theta}({\bf y})) &= \frac{1}{M} \sum_{j=1}^{M} \Bigg\{ \left\| {\bf y}^{(j)} - h_{\theta}({\bf y}^{(j)}) \right\|^2 - N \sigma^2 \nonumber \\
&\quad + \frac{2 \sigma^2}{\epsilon} (\tilde{\bf b}^{(j)})^T \left( h_{\theta}({\bf y}^{(j)} + \epsilon \tilde{\bf b}^{(j)}) - h_{\theta}({\bf y}^{(j)}) \right) \Bigg\} \tag{6}
\end{align}

where \( M \) denotes the batch size, \( \mathbf{y}^{(j)} \) symbolizes the j-th noisy image in the batch, \( N \) the dimensionality of the data, and \( \tilde{\mathbf{b}}^{(j)} \) an auxiliary Gaussian perturbation vector independent of the noise inherent in \( \mathbf{y} \). The terms within the brackets incorporate a direct error term, a noise variance adjustment, and a Monte Carlo estimation of the divergence term, providing a rigorous yet tractable surrogate for the MSE, which is typically inaccessible in unsupervised settings. This methodology not only streamlines the training process by removing the necessity for clean data but also enhances the neural network's ability to generalize from noisy inputs by effectively learning to denoise through a self-supervised learning framework.

\section{Methods}
In the following discourse, we delineate the methodologies employed for the unsupervised training of deep learning architectures dedicated to image denoising, predicated on Stein's Unbiased Risk Estimator (SURE) and its derivatives. These architectures operate under the assumption of Gaussian-distributed noise contaminants, as formalized in the initial exposition. This discourse extends to a critical reassessment of the Noise2Noise framework\cite{Jaak2018N2N}, SURE, and its enhanced variant, eSURE. Additionally, we integrate the Poisson Unbiased Risk Estimator (PURE) and its augmented form, ePURE, tailored for the denoising of Poisson-afflicted MRI images.

\subsection{Revisiting Noise2Noise}
The Noise2Noise paradigm, initially propounded for the training of Deep Neural Networks (DNNs) solely with noise-afflicted images, necessitates dual noise realizations per image\cite{Jaak2018N2N}. Despite its empirical validation, theoretical underpinnings regarding the independence or non-correlation of these noise realizations remain elusive. Under the assumption that the triplet \((x, y, z)\) adheres to a joint distribution, and considering the zero-mean vector condition of the noise differentials \(y - x\) and \(z - x\), the Mean Squared Error (MSE) for an infinite dataset can be described as:
\[
\mathbb{E}_{(\mathbf{x}, \mathbf{y})} \left\{ \|\mathbf{x} - \mathbf{h}_{\theta}(\mathbf{y})\|^2 \right\} = \mathbb{E}_{\mathbf{x}} \left[ \mathbb{E}_{(\mathbf{y}, \mathbf{x}) | \mathbf{x}} \left\{ \|\mathbf{x} - \mathbf{z} + \mathbf{z} - \mathbf{h}_{\theta}(\mathbf{y})\|^2 \big| \mathbf{x} \right\} \right]
\]
\[
= \mathbb{E}_{\mathbf{x}} \left[ \mathbb{E}_{(\mathbf{y}, \mathbf{x}) | \mathbf{x}} \left\{ \|\mathbf{t} - \mathbf{h}_{\theta}(\mathbf{y})\|^2 + 2 (\mathbf{z} - \mathbf{x})^T \mathbf{h}_{\theta}(\mathbf{y}) \big| \mathbf{x} \right\} \right] + \text{const.} \tag{7}
\] 
For a stationary \(x\), provided \(y\) and \((z - x)\) are uncorrelated or independently distributed such that \((z - x)\) maintains a zero mean vector, the resultant Noise2Noise loss function becomes:
\[
\mathbb{E}_{(\mathbf{x},\mathbf{y},\mathbf{x})}\|\mathbf{z}-h_{\theta}(\mathbf{y})\|^{2}. \tag{8}
\]
This formulation elucidates that optimal parameters \(\theta\) of a denoiser, trained under this paradigm, should ideally converge to the solution equivalent to that obtained through MSE-based training with pristine ground truth. Notably, Noise2Noise has demonstrated superior performance across diverse image restoration tasks, particularly in Gaussian noise abatement scenarios, provided the existence of two noisy realizations per ground truth image.

This analytical framework also posits that training denoisers with mildly noise-corrupted ground truth images (notated as \(x^\sim\)) augmented with additional synthetic noise might not attain efficacy comparable to training with dual independent noise realizations or pristine ground truth supplemented with synthetic noise. This hypothesis stems from the potentially significant non-zero terms of \(\mathbb{E}_{(\mathbf{x},\mathbf{y},\mathbf{z})}\left\{(\mathbf{z}-\mathbf{x})^{T} h_{\theta} (\mathbf{y})\right\}\).

\subsection{Extended Stein's Unbiased Risk Estimator and Monte-Carlo SURE}
The original Stein's Unbiased Risk Estimator (SURE) inherently handles single noise realizations per image, limiting its applicability in scenarios where multiple independent noise samples are available or required. This limitation prompts the extension of SURE to eSURE (Extended SURE), to encompass dual noisy instances per image, drawing upon the conceptual framework of Noise2Noise (N2N).

In this extension, suppose $\mathbf{y}_1 \sim \mathcal{N}(\mathbf{x}, \sigma_{\mathbf{y}_1}^2 \mathbf{I})$ represents a ground-truth image contaminated by Gaussian noise, and $\mathbf{z} \sim \mathcal{N}(\mathbf{0}, \sigma_{\mathbf{z}}^2 \mathbf{I})$ signifies additional additive white Gaussian noise (AWGN). This configuration yields a noisy observation $\mathbf{y}_2 = \mathbf{y}_1 + \mathbf{z}$, which conforms to the distribution $\mathcal{N}(\mathbf{x}, (\sigma_{\mathbf{y}_1}^2 + \sigma_{\mathbf{z}}^2) \mathbf{I})$. The eSURE is then delineated by the following unbiased estimator of the mean squared error (MSE):
\[
\mathbb{E}_{\mathbf{y}_2}\left\{\frac{1}{N}\|\mathbf{x} - h_\theta(\mathbf{y}_2)\|^2\right\} = \mathbb{E}_{\mathbf{y}_2}\left\{\gamma(h_\theta(\mathbf{y}_2), \mathbf{y}_1)\right\},
\]
where $\gamma(h_\theta(\mathbf{y}_2), \mathbf{y}_1)$ is defined as:
\[
\gamma(h_\theta(\mathbf{y}_2), \mathbf{y}_1) = \frac{1}{N} \|\mathbf{y}_1 - h_\theta(\mathbf{y}_2)\|^2 - \sigma_{\mathbf{y}_1}^2 + \frac{2 \sigma_{\mathbf{y}_1}^2}{N} \sum_{i=1}^N \frac{\partial h_i(\mathbf{y}_2)}{\partial (\mathbf{y}_2)_i}. \tag{9}
\]

The MC-SURE method facilitates the estimation of this divergence term, which is often computationally challenging to compute directly. By employing a Monte Carlo approximation:
\[
\mathbb{E}_{\tilde{\mathbf{b}}} \left\{\tilde{\mathbf{b}}^T \left(\frac{h_\theta(\mathbf{y}_2 + \epsilon \tilde{\mathbf{b}}) - h_\theta(\mathbf{y}_2)}{\epsilon}\right)\right\}, \tag{10}
\]
where $\tilde{\mathbf{b}} \sim \mathcal{N}(0, \mathbf{I})$ is an auxiliary Gaussian vector, independent of $\mathbf{y}_2$. This stochastic approximation enables an efficient and robust estimation of the gradient terms critical for the optimization of denoising networks.

These frameworks (eSURE and MC-SURE) leverage correlated noise characteristics, bridging the gap between theoretical estimators and practical denoising applications. They extend the utility of Noise2Noise by enhancing its adaptability to scenarios where dual noisy samples are derived from correlated, rather than strictly independent, noise distributions.

\section{Experiment and Result}

\subsection{Dataset and Preprocessing}
For our experiments, we utilized a dataset comprising fully-sampled 3T knee MRI scans from 22 subjects (11 males and 11 females), as referenced in \cite{knee2013}. The dataset encompasses high-resolution, complex-valued volumes for each subject, segmented into 320$\times$320$\times$256 matrices, subsequently sliced into 320$\times$256 axial planes. MRI acquisition was conducted using a 3T whole-body scanner, employing a sagittal 3D FSE CUBE sequence with proton density weighting and fat saturation. The raw k-space data was meticulously preserved for authenticity in subsequent processing.

Each knee was strategically positioned within an 8-channel HD knee coil, ensuring alignment from anterior to posterior with a precision tolerance of $\pm$10 degrees relative to the isocenter. From these scans, we generated PNG formatted images, deriving both axial and coronal views to serve as the basis for our training and evaluation.The experimental framework was structured with a dataset of 1000 axial and 1000 coronal view images utilized for training our models. An additional set of 100 images per view category was designated for testing, ensuring a comprehensive evaluation of our denoising approach across varying perspectives.

\subsection{Experimental setup}
We have conducted two experiments to evaluate all the methods to compare. In the first experiment, we
experimentally show that eSURE efficiently utilizes given two uncorrelated realizations per image
to outperform N2N,SURE and BM3D and is a general case of Noise2Noise for i.i.d. Gaussian noise with sigma value 25. Second
experiment aimed to investigate the same claim but this time with enhanced noise gaussian noise sigma value with 50. This eSURE method was compared with BM3D \cite{Kos2007}, DnCNN trained on
MC-SURE \cite{sha2018}, Noise2Noise \cite{Jaak2018N2N} and DnCNN trained with MSE using noiseless ground truth data.
We used DnCNN [6, 8] as a deep denoising network for grayscale 3T Knee MRI\cite{knee2013} images. DnCNN
consists of 20 layers of CNN with batch normalization followed by ReLU as a non-linear function. An initial learning rate was set to 10-3, which was dropped to 10-4 after 40 epochs and batch size was 128. For our work we have used Pytorch, the experiments are conducted using NVIDIA GeForce RTX 3090 with 24GB RAM.. To determine image quality, we evaluated the peak signal-to-noise ratio (PSNR), and structural similarity index (SSIM).

\subsection{Empirical Evaluation}

\begin{figure*}[!tbp]
    \centering
    \includegraphics[width=1\textwidth]{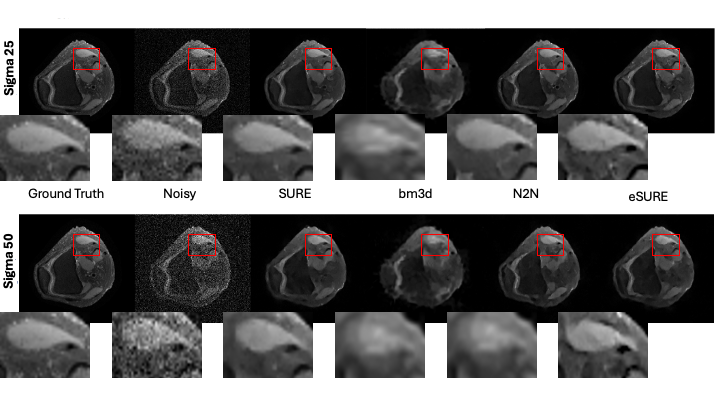}
    \caption{Denoised results of BM3D, DnCNN trained with various methods on Axial for Gaussian noise $\sigma = 25$ and $50$.}
    \label{fig:resn}
\end{figure*}

\begin{figure*}[!tbp]
    \centering
    \includegraphics[width=1\textwidth]{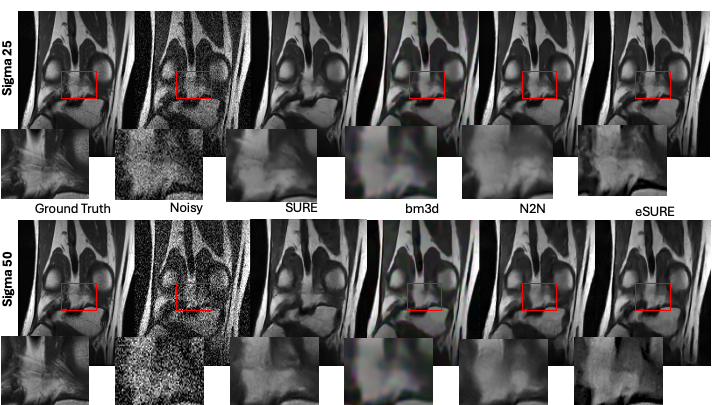}
    \caption{Denoised results of BM3D, DnCNN trained with various methods on Coronal for Gaussian noise $\sigma = 25$ and $50$.}
    \label{fig:resn1}
\end{figure*}

Table \ref{tab:mytable} delineates the comparative analysis of denoising efficacies among several state-of-the-art methodologies, including the extended Monte Carlo SURE (MC-SURE), under standardized test conditions within both axial and coronal views of 3T Knee MRI datasets. Notably, the experimental outcomes corroborate the theoretical postulates, illustrating that the eSURE framework, when deployed with dual uncorrelated noise realizations, substantiates an improvement over traditional SURE and aligns closely with Noise2Noise-induced DnCNN (DnCNN-N2N) benchmarks. Specifically, eSURE manifests a consistent supremacy over conventional SURE and DnCNN-SURE across varying noise thresholds, affirming its robustness in more realistic noise-dense medical imaging scenarios.
\begin{table}[h]
\centering
\begin{tabularx}{\textwidth}{l|XXXXXX}
    \hline
    \multicolumn{1}{c}{} & \multicolumn{5}{c}{3T Knee MRI Axial View} \\
    \hline
    Methods & BM3D & DnCNN-SURE & DnCNN-SURE* & DnCNN-N2N & DnCNN-eSURE & DnCNN-MSE \\
    \hline
    $\sigma$ = 25 & 29.10 & 31.56 & 29.00 & \textbf{33.96} & \textbf{33.96} & 29.20 \\
    $\sigma$ = 50 & 27.75 & 31.55 & 26.07 & 31.53 & \textbf{31.63} & 26.22 \\
    \hline
    \multicolumn{1}{c}{} & \multicolumn{5}{c}{3T Knee MRI Coronal View} \\
    \hline
    Methods & BM3D & DnCNN-SURE & DnCNN-SURE* & DnCNN-N2N & DnCNN-eSURE & DnCNN-MSE \\
    \hline
    $\sigma$ = 25 & 29.10 & 32.55 & 29.00 & 32.46 & \textbf{32.56} & 30.82 \\
    $\sigma$ = 50 & 27.75 & 28.75 & 26.07 & 28.86 & \textbf{29.99} & 28.83 \\
    \hline
\end{tabularx}
\caption{PSNR results of blind denoisers}
\label{tab:mytable}
\end{table}
Figure \ref{fig:resn} and \ref{fig:resn1} provides a visual corroboration of the denoising capabilities where eSURE conspicuously preserves textural integrity and edge sharpness, starkly contrasting with the blurred outputs from BM3D and underperforming variants of DnCNN-SURE. The comparative visuals distinctly demonstrate the superior retention of details by eSURE, closely mirroring the performance of DnCNN-N2N across both depicted noise levels.

In summation, the empirical validations suggest that the integration of two uncorrelated noise realizations within the eSURE framework not only enhances the denoising accuracy but also ensures a closer approximation to the MSE-trained DnCNN benchmarks, thereby extending the utility of unsupervised learning paradigms in clinical diagnostic settings where access to pristine ground truths is often unfeasible.

\section{Poisson Unbiased Risk Estimator (PURE)}
The Poisson Unbiased Risk Estimator (PURE) is a statistical tool developed to facilitate the training of deep neural networks (DNNs) for image denoising, especially under conditions where only noisy data is available. Traditionally, denoising algorithms depend on pairs of noisy and clean images; however, PURE allows for the training of denoisers using only noisy observations by estimating the underlying clean image statistics. This method is particularly relevant for medical imaging and other applications where obtaining clean ground truth data is impractical.

\subsection{Extended Poisson Unbiased Risk Estimator (ePURE)}
Building on the foundation laid by PURE, the Extended Poisson Unbiased Risk Estimator (ePURE) refines its approach to better suit the demands of self-supervised learning in environments where even low-noise images are unavailable. The extended version utilizes pairs of correlated noisy images to improve the stability and performance of the risk estimator. The mathematical representation of ePURE is as follows:
\begin{equation}
\begin{split}
ePURE(\theta) = \frac{1}{M} \sum_{j=1}^{M} \bigg[ & \| h_{\theta}(y_{1}^{(j)}) - z^{(j)} \|^2 - \frac{1}{T} \sum_{i=1}^{K} z^{(j)}[i] \\
& + \frac{2}{\epsilon T} (n^{(j)} \odot z^{(j)})^T (h_{\theta}(y_{1}^{(j)} + \epsilon n^{(j)}) - h_{\theta}(y_{1}^{(j)})) \bigg]
\end{split}
\end{equation}
Here, \(z = \frac{y_1 + y_2}{2}\) represents the mean of two correlated noisy samples, which reduces the variance of the noise in the training data, thereby enhancing the denoising performance of the network.

\subsection{Experimental Results}
The ePURE methodology was rigorously tested using a deep convolutional neural network (DnCNN) trained across a range of synthetic noise levels to mimic practical imaging scenarios(see fig. 3). This testing validated ePURE's robustness in reducing noise while preserving the integrity of the image details, as demonstrated by the measured improvements in both PSNR and SSIM (see table 2).
\begin{figure}[H]
    \centering
    \includegraphics[width=\textwidth]{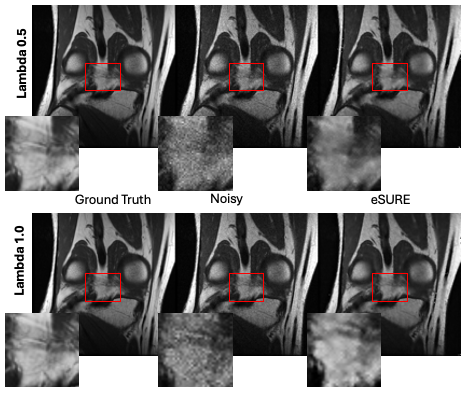}
    \caption{Visual comparison of denoising performance on MRI coronal view with Poisson noise at different intensity levels using ePURE.}
    \label{fig:ePURE_visuals}
\end{figure}
\begin{table}[h]
\centering
\begin{tabularx}{\textwidth}{l|XX}
    \hline
    \textbf{Noise Level} & \textbf{PSNR} & \textbf{SSIM} \\
    \hline
    $\lambda = 0.5$ & 31.68 & 0.934 \\
    $\lambda = 1.0$ & 32.45 & 0.947 \\
    \hline
\end{tabularx}
\caption{PSNR and SSIM results for ePURE on 3T Knee MRI Coronal View}
\label{tab:ePURE_results}
\end{table}

\section{CONCLUSION}
In this work, we have rigorously examined the effectiveness of the Noise2Noise framework and significantly advanced the state-of-the-art by adapting the Stein's Unbiased Risk Estimator (SURE) to the needs of medical imaging, specifically MRI. Our extensions, eSURE and ePURE, are tailored to exploit the noisy data inherent in medical imaging scenarios, where pristine, noise-free ground truth is rarely available. Our implementations have successfully demonstrated that eSURE efficiently harnesses information from two uncorrelated noisy images to outperform the conventional SURE approach. This capability is not just theoretical but has shown practical superiority in direct comparisons against established methods like BM3D and the original SURE, particularly in handling realistic noise models that are typical in clinical MRI data.

Moreover, our another adaptation, ePURE, has been specifically crafted to address the challenges posed by Poisson noise, which is prevalent in various medical imaging techniques. This method has proven to be comparable to existing solutions by providing high-quality denoising that hopefully facilitates better diagnostic imaging. The implications of our work are profound for clinical practice. By enhancing the quality of denoised images without requiring clean ground truth, we enable more accurate diagnostics and potentially broader applications in medical research. This advancement represents a significant stride forward in medical imaging technology, showcasing how cutting-edge machine learning techniques can be effectively applied to real-world challenges in healthcare.

\end{document}